\documentstyle[12pt,epsf]{article}

\def\be{\begin{equation}}
\def\ee{\end{equation}}
\def\bea{\begin{eqnarray}}
\def\eea{\end{eqnarray}}
\def\nn{\nonumber}

\newcommand{\Section}[1]{\section{#1}\setcounter{equation}{0}}

\begin{document}

\begin{flushright}
hep-th/0403174 \\
IPM/P-2004/011
\end{flushright}

\pagestyle{plain}

\def\e{{\rm e}}
\def\cs{\frac{1}{(2\pi\alpha')^2}}
\def\CV{{\cal{V}}}
\def\haf{{\frac{1}{2}}}
\def\tr{{\rm Tr}}
\def\"{\prime\prime}
\def\p{\partial}
\def\tphi{\tilde{\phi}}
\def\ttheta{\tilde{\theta}}
\def\a{\alpha}
\def\b{\beta}
\def\la{\lambda}
\def\barla{\bar{\lambda}}
\def\ep{\epsilon}
\def\hj{\hat j}
\def\hn{\hat n}
\def\bz{\bar{z}}
\def\zk{{\bf{Z}}_k}
\def\h1{\hspace{1cm}}
\def\dd{\Delta_{[N+2k] \times [2k]}}
\def\ddbar{\bar{\Delta}_{[2k] \times [N+2k]}}
\def\u{U_{[N+2k] \times [N]}}
\def\ubar{\bar{U}_{[N] \times [N+2k]}}
\def\goes{\rightarrow}
\def\goal{\alpha'\rightarrow 0}
\def\ads2{AdS_2 \times S^2}
\def\ola{\overline {\lambda}}
\def\oep{\overline {\epsilon}}

\vspace{3cm}

\begin{center}
{\LARGE {${\cal N}=1/2$ Super Yang-Mills Theory on Euclidean  $AdS_2 \times S^2$  }}
\end{center}

\vspace{.15cm}

\begin{center}

\vspace*{15mm} \vspace*{1mm} {Ali Imaanpur$^{a,b}$ and Shahrokh
Parvizi$^a$}

\vspace*{1cm}

{\it $^a$ Institute for Studies in Theoretical Physics and Mathematics
(IPM)\\
P.O. Box 19395-5531, Tehran, Iran\\
\vspace*{1mm}
$^b$ Department of Physics, School of Sciences \\
Tarbiat Modares University, P.O. Box 14155-4838, Tehran, Iran}\\
\vskip .1 in

{\sl aimaanpu, parvizi@theory.ipm.ac.ir}
\vspace{1cm}
\end{center}
\vskip .02 in
\begin{abstract}
We study D-branes in the background of Euclidean $AdS_2 \times S^2$ with a
graviphoton field turned on. As the background is not Ricci flat,  the graviphoton
field must have both self-dual and antiself-dual parts. This, in general, will break
all the supersymmetries on the brane. However, we show that there exists a limit for
which one can restore half of the supersymmetries. Further, we show that in this limit,
the ${\cal N}=1/2$ SYM Lagrangian on flat space can be lifted on to the Euclidean
$AdS_2 \times S^2$ preserving the same amount of supersymmetries as in the flat case.
We observe that without the $C$-dependent terms present in the action
this lift is not possible.
\end{abstract}

\vspace{3cm}
%
\newpage
\Section{Introduction}

Recent studies on string theory in the background of a graviphoton field
have revealed new structures on the worldvolume of the corresponding
D-branes. In fact, it is found that in such a background the odd coordinates
of the superspace turn out to be nonanticommuting. There are, though, two
different approaches
to the problem. Either, one could insist on preserving the whole ${\cal N}=1$
supersymmetry, as in the work of Ooguri-Vafa \cite{CVAFA}. Or, as in
\cite{SEI, SEI2}, one could assume that the anticommutation relations between
the odd coordinates on superspace survive the field theory limit. In the latter
case, however, one loses half of the
supersymmetries, and so it is called ${\cal N}=1/2$ super Yang-Mills theory.
Superspace deformations of this kind have been studied in some earlier works
\cite{CAS, SCH, BOU, FERR, KLEMM}. But the fact that it arises in a natural way from
string theory was the consequence of recent works \cite{GRASSI, CVAFA, SEI}.
Different aspects of ${\cal N}=1/2$ supersymmetric models have further been studied in
\cite{REY1}-\cite{GHOD}. While the instanton solutions and some nonperturbative effects
\cite{SOH}-\cite{BILLO}, along with the generalizations to ${\cal N}=2$ , and other
interesting features of noncommutative superspace have been explored in
\cite{ARAKI}-\cite{WOLF}.

So far, the study of D-branes in the presence of a graviphoton field has been
restricted to flat space times. Roughly speaking, the supergravity field equations
imply that for having a flat background, one has to choose a graviphoton
field which has a zero energy momentum tensor. In Euclidean signature, on the other
hand, one way of getting a zero energy momentum tensor is to choose the graviphoton
field to be (anti)self-dual. For a typical graviphoton field, however, the
energy momentum tensor is not necessarily vanishing. And hence, the
supergravity equations would imply that the spacetime cannot be flat. In Lorentzian 
signature, a well known example of this type is $\ads2$ together with a 
``self-dual'' graviphoton field. In this article, we will be studying the Euclidean
version of this solution.

Upon rotation to Euclidean space, the graviphoton field will have both self-dual
and antiself-dual parts. From the standard arguments in string theory it then
follows that the anticommutation relations for both right-handed and left-handed
odd coordinates on the D-brane get deformed. Furthermore, this also affects the
supersymmetry algebra breaking
all the supercharges on the brane. In this article, we take a scaling limit where the
self-dual part becomes very large, and at the same time the antiself-dual part goes to
zero. The limiting process, however, leaves the energy momentum tensor unaffected
and so  the $\ads2$ background is left unchanged. As a result of this limiting half
of the supersymmetries can be restored, just as in the flat case.
Having the ${\cal N}=1/2$ supersymmetry algebra, we go on to define the
corresponding Lagrangian on the $\ads2$ background. In doing so, we will make
two interesting observations. Firstly, we observe that for having a
supersymmetric Lagrangian in this background one actually does need the extra
$C$-dependent terms present in the Seiberg action. In other words, it is not
possible to have a pure ${\cal N}=1$ SYM action on $\ads2$. Secondly, even for
having an ${\cal N}=1/2$ supersymmetric Lagrangian in this background we need
to modify the supersymmetry transformations. However, we show that the
modified supersymmetry transformations continue to be a realization of
${\cal N}=1/2$ supersymmetry algebra.

At the end, we comment on the plane wave limit of $\ads2$ in the Lorentzian
signature. In this limit the deformation parameter $C$ will have a zero
determinant, which allows rotation to a frame where half of the odd coordinates
become anticommuting. However, making the corresponding supercharges
anticommute is not straightforward. In fact, we find a linear combination of
supercharges that anticommute, but then they will not act as derivations.

\Section{String Theory on $\ads2$ with RR Graviphoton Background}

Consider type-II string theory compactified on a 6-dimensional
Calabi-Yau $M$. Besides the NS-NS fields, we are also
considering form fields which come from the compactification of
RR form fields on the Calabi-Yau cycles. Among these
fields there is a one form field, the graviphoton field, which
can be obtained from wrapping RR-forms on cycles of $M$. In pure spinor
notation the field strength of this graviphoton can be shown as follows:
\bea
P^{\a\hat\b}&=&\haf P^{\mu\nu} (\sigma_{\mu\nu})^{\a\hat\b}  \\
P^{\dot\a\hat{\dot\b}}&=&\haf P^{\mu\nu}
(\bar\sigma_{\mu\nu})^{\dot\a\hat{\dot\b}} \, ,
\eea
where by construction $P^{\a\hat\b}$ and $P^{\dot\a\hat{\dot\b}}$ are self-dual
and antiself-dual parts of the graviphoton field, respectively. Here
$(\a, \dot\a, \hat\a, \hat{\dot\a} )$ show the left-handed and right-handed
indices of ${\cal N}=2$ fermions.

In Euclidean signature, the energy momentum tensor of a self-dual
(antiself-dual) graviphoton field  is zero, and thus there is no back reaction
on the geometry. However, if we consider a typical graviphoton field with both
self-dual and antiself-dual parts, there will be a back reaction. In this
article, we will be interested in the specific example of Euclidean $\ads2$.
We are going to examine the branes in this background, and in the next section
write down their effective SYM Lagrangian. In  Lorentzian signature, this
background together with a ``self-dual'' graviphoton field is a well-known
supergravity solution, on which the string theory is already studied
\cite{Zhou, 9907200}.

Let us start with the string action on flat space, and in the hybrid or
pure spinor formalism \cite{9907200}:
\bea
S_{flat}= \frac{1}{\a'}\int d^2z \left( \haf {\p}_{\bar z}x^\mu\p x_\mu +
p_\a {\p}_{\bar z}\theta^\a + \bar{p}_{\dot\a} {\p}_{\bar z}\bar{\theta}^{\dot\a} + {p}_{\hat\a} \p{\theta}^{\hat\a} +  
\bar{p}_{\hat{\dot{\a}}} \p \bar{\theta}^{\hat{\dot{\a}}} \right)
\eea
where $p$'s are the momentum conjugate to $\theta$'s. Following
\cite{9907200}, we introduce a new set of variables (and similar definitions
for $\hat{d},\hat{\bar{d}},\hat{q}$ and $\hat{\bar{q}}$):
\bea
y^\mu &=& x^\mu+i \theta^\a \sigma^\mu_{\a\dot\a}\bar{\theta}^{\dot\a}+
i {\theta}^{\hat{\a}} \sigma^\mu_{\a\dot\a}{\bar{\theta}}^{\hat{\dot{\a}}}  \nn\\
\bar{q}_{\dot\a}&=&\bar{p}_{\dot\a}-i \theta^\a \sigma^\mu_{\a\dot\a}\p_{z} x_\mu
- \theta\theta \p_{z} \bar{\theta}_{\dot\a}+\haf \bar{\theta}_{\dot\a}\p_{z}
(\theta\theta)  \nn \\
d_{\a}&=&-{p}_{\a}-i \sigma^\mu_{\a\dot\a}\bar{\theta}^{\dot\a}\p_{z} x_\mu
+ \haf \overline{\theta\theta} \p_{z} \theta_{\a}-\frac{3}{2} \p_{z} (\theta_{\a}
\overline{\theta\theta})  \nn \\
q_{\a}&=&-p_{\a}+i \sigma^\mu_{\a\dot\a}\bar{\theta}^{\dot\a}\p_{z} x_\mu
- \overline{\theta\theta} \p_{z} \theta_{\a}+\haf \theta_{\a}\p_{z}
(\overline{\theta\theta})  \nn \\
\bar{d}_{\dot\a}&=&\bar{p}_{\dot\a}+i \theta^\a \sigma^\mu_{\a\dot\a}\p_{z} x_\mu
+\haf \theta\theta \p_{z} \bar{\theta}_{\dot\a}-\frac{3}{2}
\p_{z}( \bar{\theta}_{\dot\a} \theta\theta)\, .
\eea
By adding the vertex operator of the graviphoton field, we find the
following action:
\bea
S_{flat}= \frac{1}{\a'}\int d^2z \left( \haf \p_{\bar{z}}y^\mu\p y_\mu -
d_\a \p_{\bar{z}}\theta^\a + \bar{q}_{\dot{\a}} \p_{\bar{z}}\bar{\theta}^{\dot{\a}} -{d}_{\hat{\a}} \p{\theta}^{\hat{\a}} +  \bar{q}_{\hat{\dot{\a}}}
\p{\bar{\theta}}^{\hat{\dot{\a}}} \right. \nn\\
 \left. + \a' P^{\a\hat{\b}} d_\a {d}_{\hat{\b}}
+\a' \bar{P}^{\dot{\a}\hat{\dot{\b}}}\bar{d}_{\dot{\a}}\bar{d}_{\hat{\dot{\b}}} \right)\, .
\eea

It is now straightforward to covariantize the above action, and define it on
$\ads2$ background, as follows
\bea
S=&&\frac{1}{\a'}\int \left[\haf \Pi^c_z \Pi_{\bar{z}c}
+ d_\a \Pi_{\bar{z}}^{\a} +\bar{d}_{\dot\a} \Pi_{\bar{z}}^{\dot\a}
+ d_{\hat\a} \Pi_{\bar{z}}^{\hat\a} +
\bar{d}_{\hat{\dot\a}} \Pi_{\bar{z}}^{\hat{\dot\a}} \right.\nn\\
&&\left.\hspace{20mm}
+ \a' d_\a P^{\a\hat\b}d_{\hat\b}+ \a' \bar{d}_{\dot\a}
\bar{P}^{\dot\a\hat{\dot\b}}\bar{d}_{\hat{\dot\b}}
 \right]\, ,
\eea
where we have defined $\Pi_{j}^{A}\equiv E_M^{A} \p_j Z^M$, with $E_M^{A}$
representing the supervierbein and
$Z^M=(x^\mu, \theta^\gamma, \bar{\theta}^{\dot\gamma}, \theta^{\hat{\gamma}},
\bar{\theta}^{\hat{\dot\gamma}})$. The index $A$ indicates the tangent
superspace indices $(c, \a, \dot\a, \hat{\a}, \hat{\dot\a})$. The lowest
component of $E_\mu^c$ is the vierbein, and the lowest components of $E_M^\a$
and $E_M^{\hat{\a}}$ are the gravitini which are set to zero in our
background (no gravitino). The equations of motion for $d_\a$ and
$d_{\hat\a}$ fields read,
\bea \label{eomd}
\a' P^{\a\hat\b}d_{\hat\b}+\Pi_{\bar{z}}^{\a} &=& 0\, ,  \nn\\
\a' P^{\a\hat\b}d_{\b}-\Pi_{z}^{\hat{\a}} &=& 0\, .
\eea
Using these equations of motions, we can integrate out the corresponding
fields (as well as  $\bar{d}_{\dot\a}$ and $\bar{d}_{\hat{\dot\a}}$) to
find the following action:
\bea
S=\frac{1}{\a'}\int dz d\bar{z} \left[\haf \Pi^c_z \Pi_{\bar{z}c}
+ \frac{1}{\a'}P_{\hat{\a}\b}\Pi_{\bar{z}}^{\hat\a}  \Pi_{z}^{\b}
+ \frac{1}{\a'}P_{\hat{\dot\a}{\dot\b}}\Pi_{\bar{z}}^{\hat{\dot\a}}
\Pi_{z}^{\dot\b}
 \right]\, ,\label{A}
\eea
where $P_{\hat{\a}\b}$ and $\bar{P}_{\hat{\dot\a}{\dot\b}}$ are the
inverses of $P^{\hat{\a}\b}$ and $\bar{P}^{\hat{\dot\a}{\dot\b}}$,
respectively.

In the present set up, D-branes can be introduced by a set of consistent
open string boundary conditions as follows
\bea \label{bc}
\theta_\a=\theta_{\hat\a} ,\;\;\;\;\; d_\a=d_{\hat\a}\, .
\eea
Putting the second equation above in (\ref{eomd}) implies that
\bea
\Pi_{\bar{z}}^{\a}=-\Pi_{z}^{\hat{\a}}\, .\label{PI}
\eea
To see the consequence of the above equation, let us first introduce
the explicit forms of $\Pi_{j}^{A}$'s.
Since the background is fixed, in the expression for the supervierbeins we set
the gravitino field to zero, i.e., $\psi_\mu^{\a}=\psi_\mu^{\hat{\a}}=0$, which therefore results to
\bea
\Pi_{z}^{\hat{\a}}&=& \p_z \theta^{\hat\a} \, ,   \nn\\
\Pi_{\bar{z}}^{\a}&=& \p_{\bar{z}} \theta^{\a} \, ,\nn \\
\Pi_z^c &=& e^c_\mu \p_zx^\mu + i \p_z\theta^\a \sigma^\mu_{\a\dot\a}\bar{\theta}^{\dot\a}+
i \theta^\a \sigma^\mu_{\a\dot\a}\p_z\bar{\theta}^{\dot\a}+
i \p_z{\theta}^{\hat{\a}} \sigma^\mu_{\a\dot{\a}}\bar{\theta}^{\hat{\dot{\a}}}+
i {\theta}^{\hat{\a}} \sigma^\mu_{\a\dot{\a}}\p_z\bar{\theta}^{\hat{\dot{\a}}}\nn \\
&=& e_\mu^c \p_zy^\mu \, , \nn\\
\Pi_{\bar {z}}^c &=& e^c_\mu \p_{\bar {z}}x^\mu + i \p_{\bar{z}}\theta^\a \sigma^\mu_{\a\dot\a}\bar{\theta}^{\dot\a}+
i \theta^\a \sigma^\mu_{\a\dot\a}\p_{\bar z}\bar{\theta}^{\dot\a}+
i \p_{\bar {z}}{\theta}^{\hat{\a}} \sigma^\mu_{\a\dot{\a}}\bar{\theta}^{\hat{\dot{\a}}}+
i {\theta}^{\hat{\a}} \sigma^\mu_{\a\dot{\a}}\p_{\bar z}\bar{\theta}^{\hat{\dot{\a}}}\nn \\
&=& e_\mu^c \p_{\bar z}y^\mu \, .\label{PK} 
\eea
Further, equation (\ref{PI}) now implies the
boundary conditions on the odd coordinates
\bea
\p_z \theta^{\hat{\a}}=-\p_{\bar{z}} \theta^{\a}\, ,
\eea
which, together with the bosonic boundary conditions, defines a D-brane filling the
whole 4-dimensional space.

At this stage, by plugging (\ref{PK}) into (\ref{A}), we can write the action in terms 
of $(y^\mu, \theta^\a, \bar{\theta}^{\dot\a}, \theta^{\hat{\a}},
\bar{\theta}^{\hat{\dot\a}})$. In these coordinates, it is easy to calculate the two 
point function of $\theta$'s:
\bea
\langle \theta^\a(z,\bar{z}) \theta^\b(w,\bar{w}) \rangle =
\frac{\a'^2 P^{\a \b}}{2\pi \imath} \log \frac{\bar{z}-w}{z-\bar{w}} \;.
\eea
Following the standard argument about the non-commutativity on the brane,
we find a set of non(anti)commutative coordinates on the brane as follows:
\bea
\{\theta^\a, \theta^{\b}\} &=& \a'^2 P^{\a \b} \\
\{{\bar\theta}^{\dot\a}, {\bar\theta}^{\dot\b}\} &=& \a'^2 P^{\dot\a {\dot\b}}
\label{TH}\, .
\eea
In the field theory limit, where the low energy dynamics of D-branes is studied,  
we have to consider the limit of $\a' \goes 0$. However, the nonanticommuting
characteristics of D-branes observed above can survive the limit if
at the same time we take the limit $P^{\a \b} \goes \infty$, holding
$C^{\a \b}\equiv\a'^2 P^{\a \b}$ fixed. Therefore in such a limit, we expect
that the decoupled theory on the
worldvolume of the brane to be a deformed SYM theory on a noncommutative curved
superspace. However, a simple analogy with the supersymmetry algebra in the
flat case shows that  the above deformation will break all the supersymmetries.

Although the superspace deformation breaks all the supersymmetries, in the
following, we show that there exists a limit where we can restore half of
the supersymmetries. To see this, first note that in Euclidean signature the
energy momentum tensor of the graviphoton field reads
\be
T_{\mu\nu} = \haf C_{\mu\lambda}^{+ }C^{-}_{\rho \nu}\, g^{\lambda\rho}\, ,
\ee
where $C^+ (C^-)$ indicate the self-dual (antiself-dual) part of the
graviphoton field. Therefore, if the metric is not Ricci flat, the
supergravity equations imply that the graviphoton field must have both
self-dual and antiself-dual parts. Let us, though,
introduce a real parameter $k$ and write $T_{\mu\nu}$ differently
\be
T_{\mu\nu} = \haf \left( k C_{\mu\lambda}^{+ }\right)\left(\frac{1}{k}
C^{-}_{\rho \nu}\right)\, g^{\lambda\rho}\, .
\ee
Writing $T_{\mu\nu}$ in this way, suggests that we can define a new
graviphoton field
\be
{\tilde C}_{\mu\nu} ={\tilde C}_{\mu\nu}^{+ } + {\tilde C}^{-}_{\mu \nu}= k\,
C_{\mu\nu}^{+ } + \frac{1}{k}\, C^{-}_{\mu \nu}\, ,
\ee
with the same energy momentum tensor as before (i.e., when $k=1$).
Furthermore, this allows us to take a limit where $k\to \infty$, without
disturbing the equations of motion or the $AdS_2 \times S^2$ background.
The point, however, is that in this limit the antiself-dual part of the
graviphoton field goes to zero.\footnote{For the clarity of notation, in the
following, we drop the tilde sign of the $C$ field and keep in mind that
$C^+(C^-)$ carry a factor of $k\, (1/k)$.} Hence, recalling the definitions
\bea
C^{\a\b}=\frac{1}{2}(\sigma_{\mu\nu})^{\a\b}C^{\mu\nu +} \nn \\
C^{\dot\a\dot\b}=\frac{1}{2}({\bar\sigma}_{\mu\nu})^{\dot\a\dot\b}C^{\mu\nu -}
\nn \, ,
\eea
Eq. (\ref{TH}) implies that the usual anticommutation relation between
${\bar \theta}$ coordinates is restored in this limit, i.e. we will have
\bea
\{\theta^\a, \theta^{\b}\} &=& C^{\a \b} \\
\{{\bar\theta}^{\dot\a}, {\bar\theta}^{\dot\b}\} &=& 0\, ,
\eea
or in components we will have $C^{\a\b} \sim k/R$, and
$C^{\dot\a {\dot\b}}$ scales to zero for large $k$. In the next section,
we will further show that in the
large $k$ limit, the ${\cal N}=1/2$ SYM Lagrangian on flat Euclidean space
\cite{SEI} can be lifted on to the curved background $AdS_2 \times S^2$.

In the last part of this section, we examine the corresponding Euclidean solution
of a given solution of supergravity equations of motion in Lorentzian signature.
Specifically, it is well known that a ``self-dual''  graviphoton field
\be
C=C_{\mu\nu}\, dx^\mu\wedge dx^\nu = 2R\, (\cosh\rho\, d\tau \wedge d\rho +
\cos\theta\, d\psi \wedge d\theta) \, ,
\ee
together with a metric of $AdS_2 \times S^2$
\be
ds^2 = R^2( -\cosh^2\!\rho\, d\tau^2 + d\rho^2 + \cos^2\theta\, d\psi^2
+d\theta^2)
\ee
solve the supergravity equations. Obviously if we rotate to a Euclidean
signature by sending $\tau\to i\tau$, the rotated metric and
graviphoton field continue to solve the supergravity equations in the Euclidean
signature.\footnote{The graviphoton field becomes complex-valued upon Wick
rotation. Nevertheless, we could have instead worked with the following
real-valued field
\bea
C^+ = 2Rk\, \left( \cosh\rho\, d\tau \wedge d\rho + \cos\theta\, d\psi
\wedge d\theta\right) \nn \\
C^-=-\frac{R}{k}\, \left( \cosh\rho\, d\tau \wedge d\rho - \cos\theta\,
d\psi \wedge d\theta \right)\nn
\, ,
\eea
which also solves the supergravity field equations.}
Upon Wick rotation, the graviphoton field will have both self-dual and
antiself-dual parts
\be
C= C^+ + C^- \, ,
\ee
where
\bea
C^+ = (1+i)Rk\, \left( \cosh\rho\, d\tau \wedge d\rho + \cos\theta\, d\psi
\wedge d\theta\right) \nn \\
C^-=-(1-i)\frac{R}{2k}\, \left( \cosh\rho\, d\tau \wedge d\rho - \cos\theta\,
d\psi \wedge d\theta \right)\nn
\, ,
\eea
and we have included the $k$ factor for the limiting purposes. It is
easy to explicitly check that the above field configuration, together with
\be
ds^2 = R^2( \cosh^2\!\rho\, d\tau^2 + d\rho^2 + \cos^2\theta\, d\psi^2
d\theta^2)\, ,
\ee
is a solution to the supergravity equations in Euclidean signature.\footnote{
We note that the field $C$ is covariantly constant, and thus
satisfies both the Maxwell equations and the Bianchi identity.} Now that we
have a consistent Euclidean background solution, we can go on to discuss the
construction of supersymmetric action in this background.

\Section{${\cal N}=1/2$ SYM action on Euclidean $\ads2$}

We have chosen to study D-branes on Euclidean $\ads2$ for the following reasons.
Firstly, this background, with the corresponding graviphoton field, is a
maximally supersymmetric solution admitting maximal number of Killing
spinors. Secondly, in the Euclidean
version, the self-dual and antiself-dual parts of the graviphoton field are
independent of each other, and thus can be scaled differently.
For constructing the action, we choose the minimal coupling to the background
metric.
So basically we define the ${\cal N}=1/2$ SYM action of Seiberg \cite{SEI}
on Euclidean $\ads2$ simply by covariantizing the ordinary derivatives both in
the action and in the supersymmetry transformations. This will then reduce to the
action on flat space in the large $R$ limit. The invariance of the action
under the $Q$ supersymmetry, however, is not obvious and we are going to check it
in detail. Along the way, we make two important observations. Firstly, we observe
that this lift is not possible for pure ${\cal N}=1$ SYM action, and in fact one
does need the extra
$C$ terms in the action for having invariance under $Q$. Secondly, even in the
case of $C$-deformed theory, we need to deform the  supersymmetry variation of
$\ola$ to have a supersymmetric action. 

To begin with, let us consider the deformed action of Seiberg \cite{SEI} on a
curved D-brane which has filled the Euclidean $\ads2$ space:
\bea \label{action}
S_{(C\neq 0)}= \int \sqrt{ g}\,  d^4x \tr \left[-\haf F_{\mu\nu}F^{\mu\nu}
-2i  \la \sigma^\mu D_\mu \ola
-i  C^{\mu\nu} F_{\mu\nu}\ola\ola
+\frac{1}{4} |C|^2 (\ola\ola)^2 \right]\, ,
\eea
where $D_\mu\equiv \nabla_\mu +[A_\mu,\  ]$, and $\nabla_\mu$ is the
covariant derivative on the curved background $\ads2$. We now show that
the above action is invariant under the following deformed supersymmetry
transformations
\bea \label{trans}
\delta \la &=& \sigma^{\mu\nu}\ep (F_{\mu\nu}
+\frac{i }{2}C_{\mu\nu}{\ola\ola})     \nn\\
\delta A_\mu &=& -i  \ola \bar{\sigma}_\mu \ep   \nn\\
\delta F_{\mu\nu}&=&i  \ep ( \sigma_\nu D_\mu-\sigma_\mu D_\nu)\ola
+ i  (\nabla_\mu \ep \sigma_\nu - \nabla_\nu \ep \sigma_\mu)\ola   \nn\\
\delta \ola &=&\frac{4\bar{\ep}}{\a' } \, ,
\eea
where now, $\ep$ and ${\bar \ep}$ are the Killing spinors on Euclidean $\ads2$
satisfying the Killing equations,
\bea
\nabla_\mu \ep^\a &=& \frac{1}{\a'}C^{\a\b}\sigma_{\mu\b\dot\a}
\bar{\ep}^{\dot\a} \nn\\
\nabla_\mu \bar{\ep}_{\dot\a} &=&\frac{1}{\a'}\, {C}_{\dot\a\dot\b}
\bar{\sigma}_{\mu}^{\dot\b\a} \ep_{\a} \, .\label{K}
\eea
Note that, for the last transformation in (\ref{trans}) to make sense, only the
$U(1)$ part of the $\ola$ field is meant to transform. So if we let 
$a=1, 2, 3$ and $a=4$ denote the $SU(N)$ and $U(1)$ 
gauge indices, respectively, then by the last transformation we mean $\delta \ola^a =4\delta^{a4}\bar{\ep}/\a'$. 
Further note that, in the scaling limit of large $k$, if we choose
$\ep \sim 1$ then the above equations imply that $\bar{\ep} \sim 1/k$. It is
now straightforward to check that the action (\ref{action}) is invariant under
the transformations (\ref{trans}). Let us see this in some more details.
The variation of terms present in the action are as follows:
\bea
\delta(-\haf F_{\mu\nu}F^{\mu\nu})&=& -2i  \ep F^{\mu\nu}\sigma_\nu D_\mu
\ola - 2i  F^{\mu\nu} \nabla_\mu \ep\, \sigma_\nu \ola  \label{1} \\
\delta(-2 i  \la\sigma^\mu D_\mu\ola)&=& -2i  (\sigma^{\mu\nu}\ep)
(F_{\mu\nu}+\frac{i }{2}C_{\mu\nu}{\ola\ola})\sigma^\rho D_\rho \ola
-\frac{8i}{\a'}  \la \sigma^\mu \nabla_\mu \bar{\ep}  \label{2} \\
\delta(-i  C^{\mu\nu}F_{\mu\nu}{\ola\ola}) &=& 2\ep\, C^{\mu\nu}\sigma_\nu
D_\mu \ola ({\ola\ola})+2 C^{\mu\nu} \nabla_\mu \ep \sigma_\nu
\ola ({\ola\ola})- \frac{8i}{\a'}  C^{\mu\nu}F_{\mu\nu}(\bar{\ep}\ola) \label{3} \\
\delta(\frac{1}{4} |C|^2 ({\ola\ola})^2)&=& \frac{4}{\a'} |C|^2 ({\ola\ola})
(\bar{\ep}\ola)\, . \label{4}
\eea
Upon using the Killing spinor Eqs. (\ref{K}), we see that the two terms
on the RHS of (\ref{1}), and the first term on the RHS of (\ref{2}) can be
combined into
\be
-2i  \ep F^{\mu\nu}\sigma_\nu D_\mu
\ola - 2i  F^{\mu\nu} \nabla_\mu \ep\, \sigma_\nu \ola  -2i  F_{\mu\nu}
(\sigma^{\mu\nu}\ep)\sigma^\rho D_\rho \ola=
\frac{8i}{\a'}  C^{\mu\nu}F_{\mu\nu}(\bar{\ep}\ola)\, ,
\ee
which cancels the last term in Eq. (\ref{3}).
The second term of (\ref{2}) cancels the first term in (\ref{3})
\be
C_{\mu\nu}\sigma^{\mu\nu}\ep\, \sigma^\rho D_\rho \ola ({\ola\ola})
+2\ep\, C^{\mu\nu}\sigma_\nu
D_\mu \ola ({\ola\ola})=0 \, .
\ee
And finally, using (\ref{K}) again, the second term of Eq. (\ref{3})
cancels the one in (\ref{4}),
\be
2 C^{\mu\nu} \nabla_\mu \ep\, \sigma_\nu
\ola ({\ola\ola})+\frac{4}{\a'} |C|^2 ({\ola\ola})(\bar{\ep}\ola)=0\, .
\ee
Also we note that the last term in (\ref{2}) goes to zero in the limit
$k\to \infty$. This completes the proof of invariance of the action under $Q$
supersymmetry.

The modification we made in the transformation of $\ola$ is necessary for two
reasons. Firstly, although it is of order $1/k$ and vanishes when
$k\to \infty$, we have to keep it as it gives rise to some finite terms when
acted on, for example, the third term in the action which is of order $k$.
Secondly, this modification makes $Q^2$ vanish on shell.
On the other hand, in the limit of $k\to \infty$, one gets back the usual
${\cal N}=1/2$ algebra.

\Section{Plane wave limit of $\ads2$}

In this section, we study the plane wave limit of our supergravity setup. To
consider this limit, we return to the Lorantzian version of $\ads2$ in the
presence of graviphoton field, and take the limit for both metric and
RR field. The $\ads2$ metric reads:
\be
ds^2 = R^2( -\cosh^2\!\rho\, d\tau^2 + d\rho^2 + \cos^2\theta\, d\psi^2
+d\theta^2)\, .
\ee
We now switch to the light cone coordinates $\tilde{x}^{\pm}=(\tau\pm \psi)/2 $,
and do the following rescaling,
\bea
x^{+}&=&\tilde{x}^{+}, \;\;\; r= \rho R, \nn\\
x^{-}&=&\tilde{x}^{-}/R^2, \;\;\; y=\theta R \, .
\eea
Taking the large limit of $R$, we arrive at the plane-wave limit of the metric and the
RR-form field,
\bea
ds^2 &=& -4 dx^+ dx^- - (r^2+y^2)(dx^{+})^2 + dr^2 +dy^2   \nn\\
C &=& dx^+ \wedge dr + dx^+ \wedge dy \, .
\eea

In the vierbeins basis $e^{a}_{\mu}$, the graviphoton field can be written as,
\bea
C^{\a\b}&=&\frac{1}{2}C^{\mu\nu} (\sigma_{ab})^{\a\b} e^{a}_{\mu} e^b_\nu\, .
\eea
Substituting the corresponding vierbeins, we find
\bea
C^{\a\b}&=& \frac{1}{2}(\sigma_{20}+\sigma_{30}+\sigma_{12}+\sigma_{13})^{\a\b}
\nn\\
\bar{C}^{\dot\a\dot\b}&=& \frac{1}{2}(\sigma_{20}+\sigma_{30}+\sigma_{12}
+\sigma_{13})^{\dot\a\dot\b}\, ,\label{P}
\eea
which, as shown in section 2, give rise to the
non-(anti)commutativity relations between the odd coordinates
\bea
\{\theta^\a, \theta^{\b}\} &=& C^{\a \b}  \nn\\
\{{\bar\theta}^{\dot\a}, {\bar\theta}^{\dot\b}\} &=& \bar{C}^{\dot\a\dot\b}.
\eea
At first sight, it seems that the above anti-commutation
relations  break all the supersymmetries. However, note that in the
plane wave limit (\ref{P}) the determinant of $C^{\a\b}$
(and $\bar{C}^{\dot\a\dot\b}$) vanishes, which means that there exists a
linear combination of $\theta^\a$'s for which some of the anticommutators are
zero. Therefore, in principle, it should be possible to restore part of the 
supersymmetry algebra. In fact, it is possible to redefine the supercharges as
\bea \label{newcharge}
\hat{Q}_{1} &=& A Q_{1} + B Q_{2} \nn\\
\hat{Q}_{2} &=& B Q_{1} + A Q_{2}\, ,
\eea
with some coefficients $A$ and $B$, such that the algebra of supercharges is 
changed into a more attractive one, 
\bea
\{\hat{Q}_2,\hat{Q}_2\}=\{\hat{Q}_1,\hat{Q}_2\}&=& 0  \nn\\
\{\hat{Q}_1,\hat{Q}_1\} &\neq& 0 \, .
\eea
The above algebra can be interpreted as the Lorentzian ${\cal N}=1/2$ SUSY
algebra for which $\hat{Q}_1$ and $\bar{\hat{Q_{\dot{1}}}}$ are broken while
$\hat{Q}_2$ and $\bar{\hat{Q_{\dot{2}}}}$ are survived. Unfortunately, since $A$ and
$B$ in (\ref{newcharge}) depend on derivatives, the new charges are not linear
in derivatives, and hence not obeying the Leibnitz's rule of derivation.

\section{Summary and Conclusion}

In trying to extend the ${\cal N}=1/2$ supersymmetric theory from flat space to a 
curved space as an effect of a graviphoton field, we observed that a graviphoton 
field with both self-dual and antiself-dual parts breaks the supersymmetry 
completely. On the other hand, taking a self-dual graviphoton field
is not a solution to the supergravity equations in a curved background. However,
we introduced a limit in which we could keep the background as Euclidean $\ads2$
while the antiself-dual part of the graviphoton is approaching zero. We showed in
this limit, the $C$-deformed SYM theory, as introduced in \cite{SEI} for flat
space, can be lifted on the $\ads2$ background. Further, by a small modification of 
the SUSY transformations, we proved the ${\cal N}=1/2$ invariance of the 
theory.

\section*{Acknowledgment}

Authors are very grateful to G. Mandal and S. Trivedi
for fruitful discussions around the subject. S. P. would like to thank
the hospitality of the department of
theoretical physics in Tata Institute of Fundamental Research,
Mumbai, and also the Center for Advanced Mathematical Sciences,
Beirut, where part of this work was done. The work of S. P. was supported
by Iranian TWAS chapter Based at ISMO.




\begin{thebibliography}{99}

\bibitem{CVAFA}
H. Ooguri, and C. Vafa, {\em The C-Deformation of Gluino and Non-planar
Diagrams}, Adv.\ Theor.\ Math.\ Phys.\  {\bf 7}, 53 (2003),
[arXiv:hep-th/0302109].


\bibitem{SEI} N. Seiberg,
{\em Noncommutative Superspace, N=1/2 Supersymmetry, Field Theory and
String Theory}, JHEP 0306 (2003) 010, [arXiv:hep-th/0305248].

\bibitem{SEI2} N. Berkovits, and N. Seiberg,
{\em Superstrings in Graviphoton Background and N=1/2+3/2 Supersymmetry},
JHEP 0307 (2003) 010, [arXiv:hep-th/0306226].

\bibitem{CAS} R.~Casalbuoni, {\em Relativity And Supersymmetries},
Phys.\ Lett.\ B {\bf 62}, 49 (1976),
{\em On The Quantization Of Systems With Anticommutating Variables},
Nuovo Cim.\ A {\bf 33}, 115 (1976), {\em The Classical Mechanics For
Bose-Fermi Systems}, Nuovo Cim.\ A {\bf 33}, 389 (1976).

\bibitem{SCH}
J.H. Schwarz, and P. van Nieuwenhuizen, {\em Speculations Concerning a
Fermionic Structure of Space-time}, Lett. Nuovo Cim. 34 (1982) 21.

\bibitem{BOU}
P.~Bouwknegt, J.~G.~McCarthy and P.~van Nieuwenhuizen, {\em Fusing the
coordinates of quantum superspace}, Phys.\ Lett.\ B {\bf 394}, 82 (1997),
[arXiv:hep-th/9611067].

\bibitem{FERR}
S. Ferrara, M.A. Lledo, {\em Some Aspects of Deformations of Supersymmetric
Field Theories}, JHEP 0005 (2000) 008, [arXiv:hep-th/0002084].

\bibitem{KLEMM}
D. Klemm, S. Penati, and L. Tamassia, {\em Non(anti)commutative
Superspace}, Class. Quant. Grav. 20 (2003) 2905, [arXiv:hep-th/0104190].

\bibitem{GRASSI}
J. de Boer, P. Grassi, and P. van Nieuwenhuizen, {\em Non-commutative
superspace from string theory}, Phys.\ Lett.\ B {\bf 574}, 98 (2003),
[arXiv:hep-th/0302078].

\bibitem{REY1}
R. Britto, B. Feng, and S.J. Rey, {\em Deformed Superspace, N=1/2
Supersymmetry and (Non)Renormalization Theorems}, JHEP 0307 (2003)
067, [arXiv:hep-th/0306215].

\bibitem{REY3}
R. Britto, B. Feng, and S.J. Rey, {\em Non(anti)commutative Superspace,
UV/IR Mixing, and Open Wilson Lines}, JHEP 0308 (2003) 001, [arXiv:hep-th/0307091].

\bibitem{GRIS}
M. T. Grisaru, S. Penati, and A. Romagnoni,
{\em Two-loop Renormalization for Nonanticommutative N=1/2 Supersymmetric
WZ Model}, JHEP 0308 (2003) 003, [arXiv:hep-th/0307099].

\bibitem{BO}
R. Britto, and B. Feng, {\em N=1/2 Wess-Zumino model is renormalizable},
Phys.\ Rev.\ Lett.\  {\bf 91}, 201601 (2003), [arXiv:hep-th/0307165].

\bibitem{ROM}
A. Romagnoni, {\em Renormalizability of N=1/2 Wess-Zumino model in
superspace}, JHEP {\bf 0310}, 016 (2003),
[arXiv:hep-th/0307209].

\bibitem{REY4}
O. Lunin, and S.J. Rey,
{\em Renormalizability of Non(anti)commutative Gauge Theories with N=1/2
Supersymmetry}, JHEP {\bf 0309}, 045 (2003),
[arXiv:hep-th/0307275].

\bibitem{REY2}
D. Berenstein, and S.J. Rey, {\em Wilsonian Proof for Renormalizability of
N=1/2 Supersymmetric Field Theories}, Phys.\ Rev.\ D {\bf 68}, 121701 (2003),
[arXiv:hep-th/0308049].

\bibitem{GHOD}
M.~Alishahiha, A.~Ghodsi and N.~Sadooghi, {\em One-loop perturbative
corrections to non(anti)commutativity parameter of N = 1/2 supersymmetric
U(N) gauge theory}, arXiv:hep-th/0309037.

\bibitem{SOH}
A.~Imaanpur, {\em On Instantons and Zero Modes of N = 1/2 SYM Theory},
JHEP {\bf 0309}, 077 (2003), [arXiv:hep-th/0308171].

\bibitem{SOH2}
A.~Imaanpur, {\em Comments on Gluino Condensates in N = 1/2 SYM Theory},
JHEP {\bf 0312}, 009 (2003), [arXiv:hep-th/0311137].


\bibitem{GRASSI2}
P.~A.~Grassi, R.~Ricci and D.~Robles-Llana, {\em Instanton calculations
for N = 1/2 super Yang-Mills theory}, arXiv:hep-th/0311155.

\bibitem{BRIT}
R.~Britto, B.~Feng, O.~Lunin and S.~J.~Rey, {\em U(N) instantons on N = 1/2
superspace: Exact solution and geometry of moduli space}, arXiv:hep-th/0311275.


\bibitem{BILLO}
M.~Billo, M.~Frau, I.~Pesando and A.~Lerda, {\em N = 1/2 gauge theory and
its instanton moduli space from open strings in R-R background},
arXiv:hep-th/0402160.


\bibitem{ARAKI}
T. Araki, K. Ito, and A. Ohtsuka, {\em Supersymmetric Gauge Theories on
Noncommutative Superspace}, Phys.\ Lett.\ B {\bf 573}, 209 (2003),
[arXiv:hep-th/0307076].


\bibitem{FER}
S. Ferrara, and E. Sokatchev, {\em Non-anticommutative N=2 super-Yang-Mills
theory with singlet deformation}, Phys.\ Lett.\ B {\bf 579}, 226 (2004),
[arXiv:hep-th/0308021].


\bibitem{IVAN}
E. Ivanov, O. Lechtenfeld, and B. Zupnik, {\em Nilpotent deformations
of N=2 superspace}, JHEP {\bf 0402}, 012 (2004),
[arXiv:hep-th/0308012].


\bibitem{TERA} S. Terashima, and J. Yee, {\em Comments on Noncommutative
Superspace}, JHEP {\bf 0312}, 053 (2003),
[arXiv:hep-th/0306237].


\bibitem{REZA}
R. Abbaspur, {\em Scalar Solitons in Non(anti)commutative Superspace},
arXiv:hep-th/0308050.

\bibitem{CHA}
M. Chaichian, and A. Kobakhidze, {\em Deformed N=1 supersymmetry},
arXiv:hep-th/0307243.

\bibitem{BAR}
I.~Bars, C.~Deliduman, A.~Pasqua and B.~Zumino, {\em
Superstar in noncommutative superspace via covariant quantization of the
superparticle}, Phys.\ Rev.\ D {\bf 68}, 106006 (2003),
[arXiv:hep-th/0308107].


\bibitem{SAK}
A.~Sako and T.~Suzuki, {\em Ring structure of SUSY $*$ product and 1/2
SUSY Wess-Zumino model}, Phys.\ Lett.\ B {\bf 582}, 127 (2004),
[arXiv:hep-th/0309076].

\bibitem{KUM}
B.~Chandrasekhar and A.~Kumar, {\em D = 2, N = 2, supersymmetric theories
on non(anti)commutative superspace}, JHEP {\bf 0403}, 013 (2004), 
[arXiv:hep-th/0310137].

\bibitem{MIK}
D.~Mikulovic, {\em Seiberg-Witten map for superfields on canonically deformed 
N = 1,\\ d = 4 superspace}, JHEP {\bf 0401}, 063 (2004), [arXiv:hep-th/0310065].


\bibitem{ISO}
S. Iso, and H. Umetsu, {\em Gauge Theory on Noncommutative Supersphere from
Supermatrix Model}, arXiv:hep-th/0311005.


\bibitem{ARAKI2}
T.~Araki, K.~Ito and A.~Ohtsuka, {\em N = 2 supersymmetric U(1) gauge
theory in noncommutative harmonic superspace}, JHEP {\bf 0401}, 046 (2004),
[arXiv:hep-th/0401012].

\bibitem{WOLF}
C.~Saemann and M.~Wolf, {\em Constraint and super Yang-Mills equations on the deformed superspace $R_{\hbar}^{(4|16)}$}, arXiv:hep-th/0401147.

\bibitem{Zhou}
J.~G.~Zhou, {\em Super 0-brane and GS superstring actions on $AdS_2 \times S^2$}, 
Nucl.\ Phys.\ B {\bf 559}, 92 (1999), [arXiv:hep-th/9906013].

\bibitem{9907200}
N.~Berkovits, M.~Bershadsky, T.~Hauer, S.~Zhukov and B.~Zwiebach,
{\em Superstring theory on $AdS_2 \times S^2$ as a coset supermanifold},
Nucl.\ Phys.\ B {\bf 567}, 61 (2000), [arXiv:hep-th/9907200].



\end{thebibliography}
\end{document}